\documentclass[11pt,a4paper]{article}
\usepackage{epsfig}

\def\etal{{\it et al.}}
\def\zm{\mbox{\rm Z}^0}

\def\ctm{\cos\theta}

\def\actm{|\ctm|}

\def\tauch{\tau^- \rightarrow 3h^-2h^+ \nu_{\tau}}
\def\tauneut{\tau^- \rightarrow 3h^- 2h^+ \pi^0\nu_{\tau}}

\def\ee{\mbox{\rm e}^+\mbox{\rm e}^-}
\def\mm{\mu^+\mu^-}
\def\tt{\tau^+\tau^-}
\def\qq{{\rm q}\overline{{\rm q}}}
\def\ggee{(\ee)\; \ee}
\def\ggmm{(\ee)\; \mm}

\def\reeee{\ee \rightarrow \ee}
\def\reemm{\ee \rightarrow \mm}
\def\reeqq{\ee \rightarrow \qq}

\def\reeeeee{\ee \rightarrow \ggee}
\def\reeeemm{\ee \rightarrow \ggmm}

\def\reeeex{\ee \rightarrow (\ee)\mbox{\rm X}\ }

\def\ztautau{\zm \! \rightarrow \! \tt}

\def\numtau{196694}
\def\numtautau{98347}

\def\dedx{\mbox{\rm d}E/\mbox{\rm d}x}

\def\fh{\tau^- \rightarrow  3h^-2h^+ \nu_{\tau}}
\def\fhp{\tau^- \rightarrow  3h^-2h^+ \pi^0 \nu_{\tau}}
\def\fhpp{\tau^- \rightarrow  3h^-2h^+ 2\pi^0 \nu_{\tau}}
\def\fp{\tau^- \rightarrow  3h^-2h^+ (\pi^0)\nu_{\tau}}

\def\finc{\tau^- \rightarrow  3h^-2h^+ (\geq 0\pi^0)\nu_{\tau}}

 \def\binc{(0.119\pm 0.013\pm 0.008)\%}
 \def\bdat{149.0}
 \def\bttb{ 21.3}
 \def\bqqb{  8.7}

 \def\bfbb{0.94\pm 0.01}

 \def\bpone{(0.091\pm 0.014\pm 0.005)\%}
 \def\bptwo{(0.027\pm 0.018\pm 0.007)\%}
 \def\nc{ 96.5\pm 14.4}
 \def\nn{ 22.6\pm 14.8}
 \def\effc{0.584\pm 0.007}
 \def\effn{0.463\pm 0.011}
 \def\biac{0.941\pm 0.009}
 \def\bian{0.931\pm 0.014}
 \def\fnt{0.020\pm 0.002}

 \def\sxsab{0.004}
 
 \def\sxsad{0.000}
 \def\sxsae{0.002}
 
 \def\sxsag{0.000}
 \def\sxsat{0.005}
 \def\sxsbb{0.006}
 
 \def\sxsbd{0.003}
 \def\sxsbe{0.001}
 
 \def\sxsbg{0.003}
 \def\sxsbt{0.007}

\begin{document}
\begin{titlepage}
\begin{center}{\large   EUROPEAN LABORATORY FOR PARTICLE PHYSICS
}\end{center}\bigskip
\begin{flushright}
       CERN-EP/98-090   \\ 4th June 1998
\end{flushright}
\bigskip\bigskip\bigskip\bigskip\bigskip
\begin{center}{\Large\bf  
Measurement of Tau Branching Ratios to Five Charged Hadrons 
}\end{center}\bigskip\bigskip

\begin{center}{\LARGE The OPAL Collaboration
}\end{center}\bigskip\bigskip
\bigskip\begin{center}{\large  Abstract}\end{center}
The  branching ratios of the decay of the $\tau$ lepton to five 
charged hadrons have been measured with the OPAL detector at LEP using 
data collected between 1991 and 1995 at $\ee$ 
centre-of-mass energies close to the $\zm$ resonance.
The branching ratios are measured to be
\[      B(\tauch) = \bpone   \]
\[      B(\tauneut) = \bptwo \]
where the first error is statistical and the second systematic. 
\bigskip\bigskip\bigskip\bigskip


\bigskip\bigskip
\begin{center}{\large
(Submitted to the European Physical Journal)
}\end{center}
\end{titlepage}
\begin{center}{\Large        The OPAL Collaboration
}\end{center}\bigskip
\begin{center}{
K.\thinspace Ackerstaff$^{  8}$,
G.\thinspace Alexander$^{ 23}$,
J.\thinspace Allison$^{ 16}$,
N.\thinspace Altekamp$^{  5}$,
K.J.\thinspace Anderson$^{  9}$,
S.\thinspace Anderson$^{ 12}$,
S.\thinspace Arcelli$^{  2}$,
S.\thinspace Asai$^{ 24}$,
S.F.\thinspace Ashby$^{  1}$,
D.\thinspace Axen$^{ 29}$
% ...
}\end{center}\bigskip
\bigskip
$^{  1}$School of Physics and Astronomy, University of Birmingham,
Birmingham B15 2TT, UK
\newline
$^{  2}$Dipartimento di Fisica dell' Universit\`a di Bologna and INFN,
I-40126 Bologna, Italy
\newline
$^{  3}$Physikalisches Institut, Universit\"at Bonn,
D-53115 Bonn, Germany
\newline
$^{  4}$Department of Physics, University of California,
Riverside CA 92521, USA
\newline
$^{  5}$Cavendish Laboratory, Cambridge CB3 0HE, UK
\newline
$^{  6}$Ottawa-Carleton Institute for Physics,
Department of Physics, Carleton University,
Ottawa, Ontario K1S 5B6, Canada
\newline
$^{  7}$Centre for Research in Particle Physics,
Carleton University, Ottawa, Ontario K1S 5B6, Canada
\newline
$^{  8}$CERN, European Organisation for Particle Physics,
CH-1211 Geneva 23, Switzerland
\newline
$^{  9}$Enrico Fermi Institute and Department of Physics,
University of Chicago, Chicago IL 60637, USA
\newline
$^{ 10}$Fakult\"at f\"ur Physik, Albert Ludwigs Universit\"at,
D-79104 Freiburg, Germany
\newline
$^{ 11}$Physikalisches Institut, Universit\"at
Heidelberg, D-69120 Heidelberg, Germany
\newline
$^{ 12}$Indiana University, Department of Physics,
Swain Hall West 117, Bloomington IN 47405, USA
\newline
$^{ 13}$Queen Mary and Westfield College, University of London,
London E1 4NS, UK
\newline
$^{ 14}$Technische Hochschule Aachen, III Physikalisches Institut,
Sommerfeldstrasse 26-28, D-52056 Aachen, Germany
\newline
$^{ 15}$University College London, London WC1E 6BT, UK
\newline
$^{ 16}$Department of Physics, Schuster Laboratory, The University,
Manchester M13 9PL, UK
\newline
$^{ 17}$Department of Physics, University of Maryland,
College Park, MD 20742, USA
\newline
$^{ 18}$Laboratoire de Physique Nucl\'eaire, Universit\'e de Montr\'eal,
Montr\'eal, Quebec H3C 3J7, Canada
\newline
$^{ 19}$University of Oregon, Department of Physics, Eugene
OR 97403, USA
\newline
$^{ 20}$Rutherford Appleton Laboratory, Chilton,
Didcot, Oxfordshire OX11 0QX, UK
\newline
$^{ 22}$Department of Physics, Technion-Israel Institute of
Technology, Haifa 32000, Israel
\newline
$^{ 23}$Department of Physics and Astronomy, Tel Aviv University,
Tel Aviv 69978, Israel
\newline
$^{ 24}$International Centre for Elementary Particle Physics and
Department of Physics, University of Tokyo, Tokyo 113, and
Kobe University, Kobe 657, Japan
\newline
$^{ 25}$Institute of Physical and Environmental Sciences,
Brunel University, Uxbridge, Middlesex UB8 3PH, UK
\newline
$^{ 26}$Particle Physics Department, Weizmann Institute of Science,
Rehovot 76100, Israel
\newline
$^{ 27}$Universit\"at Hamburg/DESY, II Institut f\"ur Experimental
Physik, Notkestrasse 85, D-22607 Hamburg, Germany
\newline
$^{ 28}$University of Victoria, Department of Physics, P O Box 3055,
Victoria BC V8W 3P6, Canada
\newline
$^{ 29}$University of British Columbia, Department of Physics,
Vancouver BC V6T 1Z1, Canada
\newline
$^{ 30}$University of Alberta,  Department of Physics,
Edmonton AB T6G 2J1, Canada
\newline
$^{ 31}$Duke University, Dept of Physics,
Durham, NC 27708-0305, USA
\newline
$^{ 32}$Research Institute for Particle and Nuclear Physics,
H-1525 Budapest, P O  Box 49, Hungary
\newline
$^{ 33}$Institute of Nuclear Research,
H-4001 Debrecen, P O  Box 51, Hungary
\newline
$^{ 34}$Ludwigs-Maximilians-Universit\"at M\"unchen,
Sektion Physik, Am Coulombwall 1, D-85748 Garching, Germany
\newline
\bigskip\newline
$^{  a}$ and at TRIUMF, Vancouver, Canada V6T 2A3
\newline
$^{  b}$ and Royal Society University Research Fellow
\newline
$^{  c}$ and Institute of Nuclear Research, Debrecen, Hungary
\newline
$^{  d}$ and Department of Experimental Physics, Lajos Kossuth
University, Debrecen, Hungary
\newline
$^{  e}$ on leave of absence from the University of Freiburg
\newline


\newpage
\noindent
{\large\bf Introduction}

\noindent
Knowledge of the tau lepton properties is becoming increasingly
more precise with the large data sets available.
Measurements of the decay modes
to a single charged particle (1-prong) and three charged 
particles (3-prong)
have been made by numerous experiments with  precision
surpassing the 1\% level \cite{pdg}.
However, only a few measurements of the $\tauch$ and $\tauneut$\footnote
{Charge conjugation is implied throughout this paper. The symbol
$h^-$ is used to indicate either $\pi^-$ or $\rm K^-$.} 
branching ratios (5-prong) have been made 
\cite{aleph_5prong,cleo_5prong,argus_5prong,hrs_5prong}.
Studies of the 5-prong decay modes are important as they are
used in the determination of the mass limit on the tau neutrino 
(for example, see ref. \cite{opaltauneutrino}).
Further the branching ratios of tau decays to five and six pions
can be compared with the predictions of an isospin model
\cite{sobie,rouge}.

This paper presents a new measurement of these modes 
using the data collected between 1991 and 1995  at energies close to
the $\zm$ resonance, corresponding to an integrated luminosity of 
163 pb$^{-1}$, with the OPAL detector at LEP.
A description of the OPAL detector can be found in ref.~\cite{opaldetector}.
The performance and particle identification capabilities of the
OPAL jet chamber are described in ref.~\cite{opaltracker}.
The tau pair Monte Carlo samples used in this analysis were
generated using the KORALZ 4.0 package \cite{koralz}.
The dynamics of the tau decays were simulated with the 
TAUOLA 2.0 decay library \cite{tauola}.
A total of 830 000 tau pair Monte Carlo events were used in this
analysis.
In addition, samples of 5000 $\fh$ and 2000 $\fhp$ Monte Carlo 
events were also used.
Both the $\fh$ and $\fhp$ decays were generated with a
uniform phase space distribution.
The $\fhpp$ decay was not simulated.
The Monte Carlo events were then passed through the 
OPAL detector simulation \cite{gopal}.


\noindent
{\large\bf Event selection}

\noindent
The procedure used to select $\ztautau$ events is similar to that described
in previous OPAL publications (for example, see \cite{opal:electron}).
The $\tt$ events are characterized by a pair of back-to-back,
narrow jets with low particle multiplicity.
The two tau jets are restricted to the barrel region of the OPAL
detector by requiring that the polar angle of the two jets satisfy
$\overline{\actm} < 0.68$ in order to avoid regions of non-uniform 
calorimeter response.
Background from other two fermion events is reduced by a number
of requirements.
Multihadronic events ($\reeqq$) are significantly reduced by requiring fewer 
than eight tracks and ten electromagnetic clusters per event.
Bhabha $(\reeee$) and muon pair $(\reemm$) events are removed by
rejecting events where the total electromagnetic energy or the
scalar sum of the track momenta are close to the centre-of-mass.
Two photon ($\reeeeee$ or $\reeeemm$) events are removed by 
rejecting events in which there is little energy in the electromagnetic
calorimeter.

In the OPAL tau pair selection events are usually required to have between
two and six tracks; however, events with up to eight tracks are allowed
in the present selection in order to increase the efficiency for $\fp$ decays.
In addition, tracks are normally required to have at least one hit
in the central drift chamber at a radius of less than 75 cm from 
the beam axis.
However, for this analysis this requirement is not imposed in order
to avoid rejecting tracks that overlap at small radii.
The efficiency for selecting tau pair events is approximately 54\%
from the Monte Carlo simulation, primarily due to the requirement
that the tau jets are in the barrel region of the OPAL detector.

A total of $\numtautau$ $\tau^+\tau^-$ candidates were selected from the 
1991-1995 data set.
The number of $\reeee$, $\reemm$ and $\reeeex$ (where X is either 
$\ee$ or $\mm$) events in the tau pair sample is unaffected by
the modifications in the standard tau pair selection and the
background fraction is estimated to be $(1.24 \pm 0.09)\%$
(see ref.~\cite{opal:electron}).
The $\reeqq$ background, however, is sensitive to the requirement
on the number of tracks and was determined to be $(0.74 \pm 0.05)\%$
in the selected sample.
The $\qq$ background in the tau pair sample was determined by
comparing the data with the expectation of the Monte Carlo 
simulation~\cite{jetset}.
The $\qq$ Monte Carlo generator is found to give more low multiplicity
$\qq$ events than are observed in the data and a correction is
made to the $\qq$ Monte Carlo.
The total fraction of background is estimated to be $(2.0 \pm 0.1)\%$.  

The $\fp$ selection begins by identifying jets with five well-measured tracks
where the absolute value of the sum of the charges of the tracks must be
equal to unity.
Approximately 72\% of the $\fp$ jets in the tau pair sample 
are selected with these requirements.
A large fraction of this sample are 
$\tau^- \rightarrow 3\pi^-(\ge 1\pi^0)\nu_\tau$ decays where 
one of the photons in the final state undergoes a conversion to an $\ee$ pair.
A neural network algorithm  \cite{idncon} found 
70\% of the 5-track jets contained a photon conversion
and these jets are removed from the sample.
The rejection of jets with photon conversions decreases the efficiency 
for selecting $\fp$ jets from 72\% to 64\%.

The sample of 5-track jets also includes 
$\tau^- \rightarrow X^- {\rm K}^0_S \nu_\tau$ 
decays where the K$^0_S$ decays to  $\pi^+\pi^-$ and $X^-$ is any
number of hadrons.
The sample also includes jets where a hadron has interacted in a part 
of the detector creating secondary particles.
After the 5-track jets with photon conversions have been removed,
approximately 20\% of the remaining 5-track jets are found to have a
pair of oppositely charged tracks with a secondary vertex in the $r-\phi$ plane.
Details of the secondary vertex algorithm can be found in \cite{opal_k0_paper}.
The 5-track jets with an identified secondary vertex are rejected 
and the $\fp$ selection efficiency decreases from 64\% to 62\%.

\begin{figure}
\begin{center}
\mbox{\epsfig{file=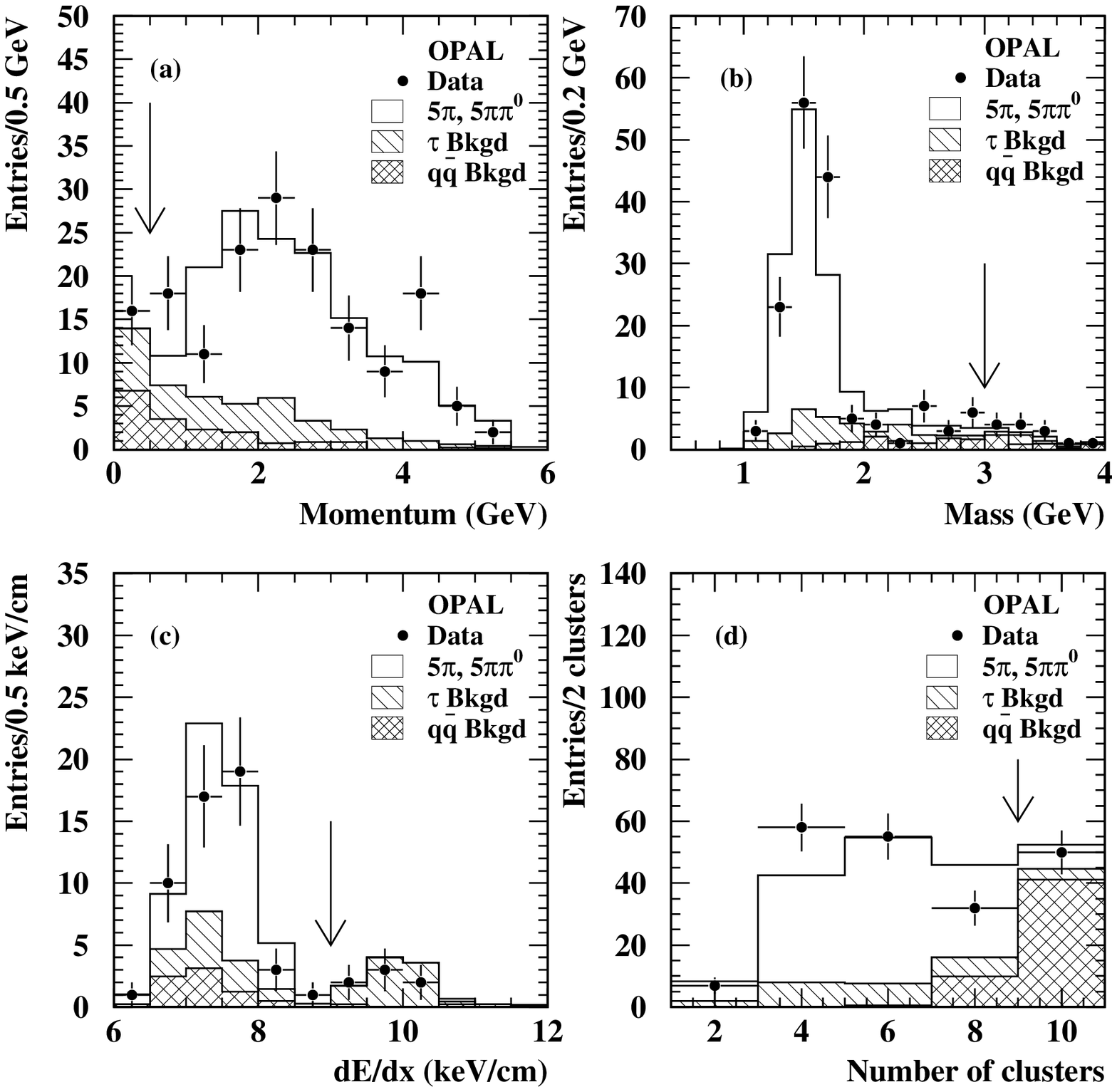,height=15cm}}
\end{center}
\caption{\label{fig:mass} 
(a) the momentum of the lowest momentum track in the jet,
(b) the mass of the five tracks,
(c) the energy loss ($\dedx$) in the central drift chamber
for tracks with $p < 2$ GeV and
(d) the number of clusters in the event.
The figures are for jets selected as $\fp$ candidates.
All $\fp$ selection criteria are applied  and the arrows indicate
the cut locations.
In all plots the world average branching ratios \cite{pdg}
are used except for the $\fh$ and $\fhp$ decay modes where the 
branching ratios obtained in this work are used.}
\end{figure}

The residual background  is reduced by requiring that
each track in the jet has a momentum ($p$) greater than 0.5 GeV 
(see Fig.~\ref{fig:mass}(a))
and that the invariant mass of the five tracks is less than 3 GeV 
assuming that the tracks are pions (see Fig.~\ref{fig:mass}(b)).
The jet is  rejected if the lowest momentum track has $p < 2$ GeV 
and $\dedx > 9$ keV/cm (see Fig.~\ref{fig:mass}(c)),
where $\dedx$ is the ionization energy deposited by the track
in the OPAL jet chamber.
The $\dedx$  on a track is only considered to be reliable if there are at 
least 20 wires out of a possible 159 wires with good $\dedx$ measurements.
Background from $\reeqq$ events is 
reduced by limiting the  number of clusters in the electromagnetic
calorimeter to a maximum of eight clusters in the event
(see Fig.~\ref{fig:mass}(d)).
These selection criteria reduce the background to approximately 20\% of the 
sample while only slightly decreasing the $\fp$ selection efficiency
from 62\% to 56\%.

A total of 152 five-track jets pass the selection.
The sample includes a background estimated to be approximately 
22 jets from other tau decays (primarily 3-prong decays) and  
10 jets from $\qq$ events.

\noindent
{\large\bf Results}

\begin{figure}
\begin{center}
\mbox{\epsfig{file=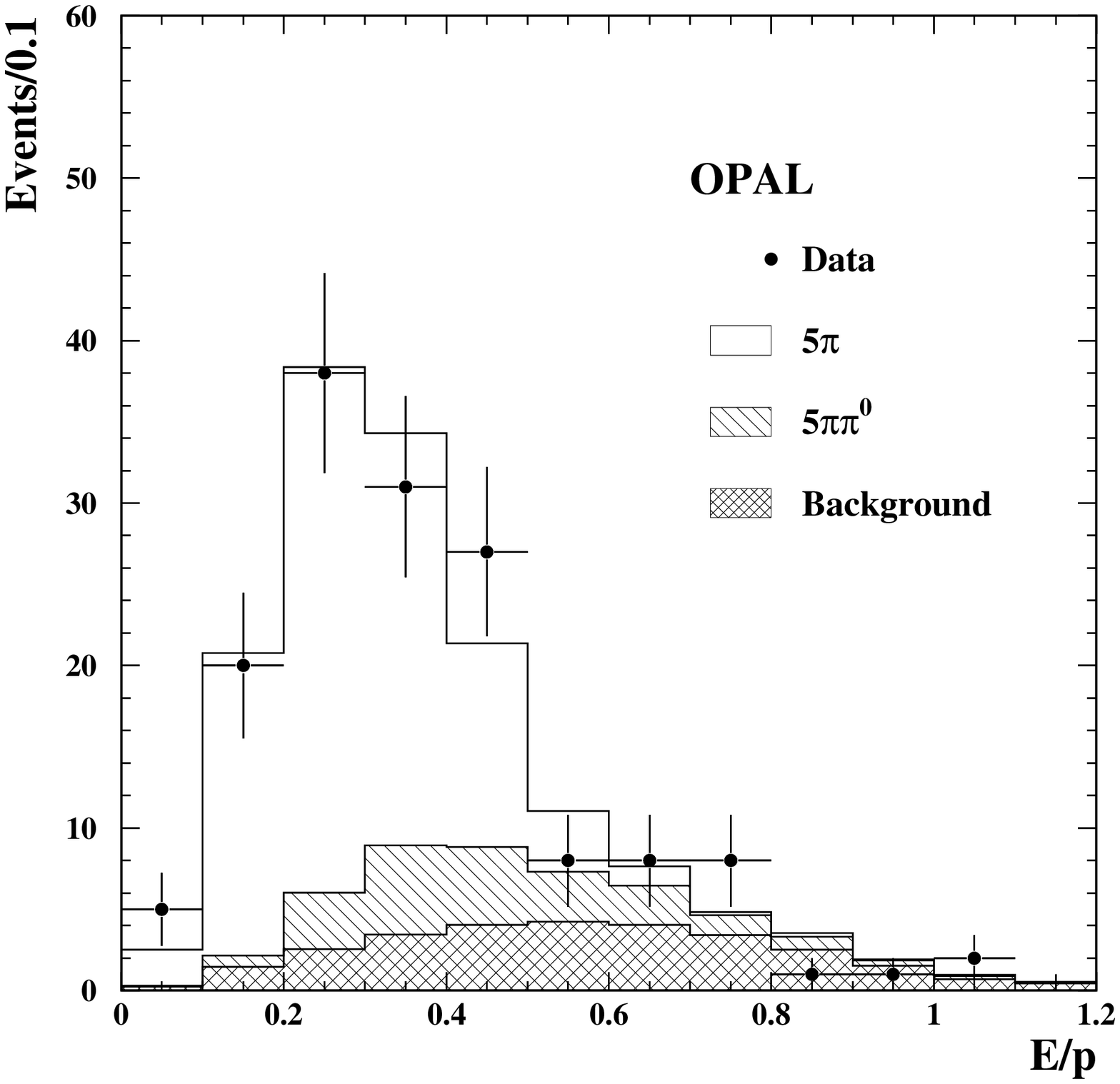,height=15cm}}
\end{center}
\caption{\label{fig:epfit}
The $E/p$ spectrum for $\fp$ jets is shown.
The histograms are the result of the fit to the data.
}
\end{figure}

\noindent
The number of $\fh$ and $\fhp$ decays is determined by 
performing a binned likelihood fit of the $E/p$ spectrum, where
$E$ is the sum of all the electromagnetic energy in the jet
and $p$ is the scalar sum of the momentum of the five tracks.
The shapes of the $E/p$ distributions are obtained from high
statistics $\fh$ and $\fhp$ Monte Carlo samples.
The $E/p$ distributions for the background are obtained from 
the  tau and $\qq$ Monte Carlo samples.
The normalization of the $\fh$, $\fhp$, tau background
and $\qq$ background $E/p$ distributions were allowed
to vary in the fit.
The uncertainties in the tau and $\qq$ backgrounds (discussed below)
were included into the likelihood fit.
The result of the fit is shown in Fig.~\ref{fig:epfit}.

The branching ratios were evaluated using 
\[ B = \frac{N}{N_{\tau} (1-f)} \; \frac{1}{\epsilon}
 \; \frac{1}{F_{\rm B}} \]
where $N$ is the number of signal events determined from the likelihood fit,
$N_{\tau}$ is the number of tau candidates,
$f$ is the non-tau background in the tau pair sample,
and $\epsilon$ is the efficiency.

The tau pair selection does not select all tau decay modes with
equal probability.
The factor $F_{\rm B}$, obtained from Monte Carlo,
corrects for the bias introduced by the tau pair selection.
In general, these factors are close to unity for most tau decay channels,
however, they are slightly less than unity for the $\fh$ and $\fhp$ 
decay channels (see Table~\ref{table:results})
due to the requirement that the tau pair events have less than 
or equal to eight tracks.
Minor variations in the tau pair selection criteria were 
found to have little impact on the value of $F_{\rm B}$.

The results of this likelihood fit are given in Table~\ref{table:results}.
The uncertainty of the background (discussed below)
was included in the likelihood fit.
The normalization factors of the tau and $\qq$ backgrounds obtained
in the fit were $0.996 \pm 0.217$ and $1.004 \pm 0.337$, respectively.
The correlation coefficient between the $\fh$ and $\fhp$ branching ratios
is found to be $-0.60$.
The systematic errors on the branching ratios include uncertainties on 
the tracking, energy resolution and  fit method  (see Table~\ref{table:syst}).

\begin{table}
\begin{center}
\begin{tabular}{lrr} \hline
                      & $\fh$     & $\fhp$      \\ \hline
Events                & $\nc$     & $\nn$       \\
Tau candidates        & \numtau   & \numtau     \\
Efficiency            & $\effc$   & $\effn$     \\
Non-tau background    & $\fnt$    & $\fnt$      \\
Bias factor           & $\biac$   & $\bian$     \\ \hline
Branching Ratio       & $\bpone$  & $\bptwo$    \\ \hline
\end{tabular}
\end{center}
\caption{\label{table:results} 
The results used to calculate the branching ratios.}
\end{table}

\begin{figure}
\begin{center}
\mbox{\epsfig{file=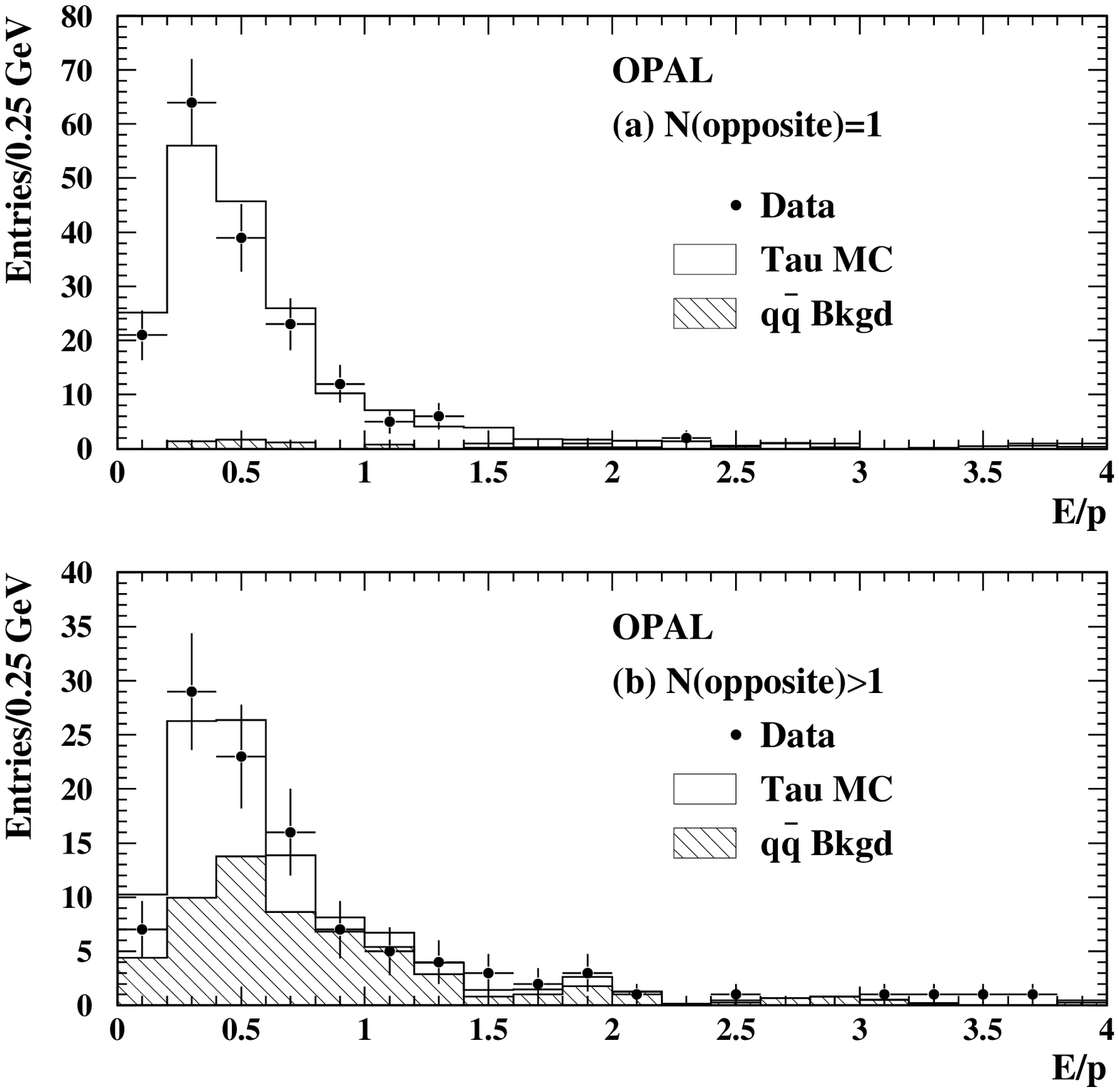,height=15cm}}
\end{center}
\caption{\label{fig:m4}
The $E/p$ spectrum for jets with four tracks is shown.
In plot (a) there is a jet on the opposite side with only one track.  
In plot (b) there is a jet on the opposite side with more than one track.
}
\end{figure}

The uncertainty in the tau and $\qq$ background includes a
statistical component based on the number of data and Monte Carlo
events in the background sample and a systematic component based
on the modelling of the tau and $\qq$ background.
The statistical uncertainty is estimated to be 20\% and 30\%
for the tau and $\qq$ backgrounds, respectively.
The systematic uncertainty is determined using a sample of events
in which one of the two jets has four tracks.
The composition of this sample (3-prong tau decays with a 
photon conversion and $\qq$ events) is very similar to the background found
in the 5-track sample.
In Fig.~\ref{fig:m4}(a)  the $E/p$ distribution of the 4-track jets is plotted
for events in which the other jet in the event
has only one track in order to enhance the tau background.
The ratio of the number of 4-track jets in the 
data versus Monte Carlo simulation is consistent with unity
and we estimate the uncertainty on the tau background to be 
approximately 10\% based on the statistical uncertainty of the
data and Monte Carlo samples.
In Fig.~\ref{fig:m4}(b), the $E/p$ distribution is plotted
when the other jet in the event has more than one track in order to
enhance the $\qq$ background.
Again, the ratio of the number of 4-track jets in the 
data versus Monte Carlo simulation is consistent with unity
and we estimate the uncertainty on the $\qq$ background to be 
approximately 20\%.
The combined statistical and systematic uncertainties
are found to be 22\% and 36\% for the tau and $\qq$
backgrounds, respectively.

The tracks in 5-prong tau decays are extremely collimated
and the results may be sensitive to the modelling of the tracks.
This was investigated by studying the distribution of the angle
between each track and its nearest neighbour.
The data were found to be well-modelled by the Monte Carlo simulation.
This comparison was repeated with a Monte Carlo  
where the track parameters were smeared  by $\pm 20\%$.
The branching ratios were evaluated using these samples and
the  change in the branching ratios was included as 
part of the tracking systematic error (see Table~\ref{table:syst}).

Possible differences in the modelling of the number of
reconstructed tracks in the data and Monte Carlo simulation were
investigated using the $\dedx$ distributions of the tracks.
Single tracks in the OPAL detector with momentum greater than 2 GeV 
should have a maximum $\dedx$ of 10 keV/cm.
Tracks with $\dedx$ substantially above 10 keV/cm are likely due to two
charged particles being reconstructed as a single track.
The fraction of jets in the 5-track sample where there is at least
one track with $\dedx > 12$ keV/cm was found to be
$0.020 \pm 0.003$ and $0.036 \pm 0.002$
in the data and Monte Carlo samples, respectively.
Similar results were obtained using the 4-track sample.
The difference between the data and Monte Carlo simulation
was accounted for by adding a 2\% uncertainty to the tracking systematic 
error (see Table~\ref{table:syst}).

Additional studies of the modelling of the tracks were made.
For example, it is possible for $\fp$ jets to appear in the 4-track sample
if a track failed one of the track quality requirements,
such as the number of hits in the tracking chamber or the impact parameter.
Branching ratios obtained using different track quality
requirements were almost unchanged and therefore no additional 
systematic error was added.

A systematic error was added to account for the uncertainty 
in the modelling of the  electromagnetic energy.
The uncertainty in the energy scale of the electromagnetic
calorimeter between the data and the Monte Carlo sample was
determined to be $\pm 0.5\%$ based on studies using tau 3-prong decays.
The uncertainty in the electromagnetic energy for the 5-prong 
tau decays was assumed to be $\pm 1\%$.
The change in the branching ratios when the energy was scaled by
$\pm 1\%$ is included in the electromagnetic energy systematic 
error (see Table~\ref{table:syst}).
In addition, the effect of smearing the electromagnetic energy on the branching
ratios was found to be negligible.

The reliability of the likelihood fit was investigated
by using the tau Monte Carlo sample to generate a many sets of
`data' and subsequently fitting it with the distributions obtained
from the high statistic signal Monte Carlo samples.
The systematic error quoted on the fit is the result of
changing the upper range of the fit (nominally $E/p=1.2$)
between $E/p=0.9$ and 1.4.

The $\fh$ and $\fhp$ branching ratios were determined assuming
that the $\fhpp$ branching ratio was negligible.
The CLEO Collaboration obtained a limit on the $\fhpp$ decay of 
$0.011\%$  \cite{cleo_5prong}, suggesting 
that this decay mode could contribute up to 10\% of the inclusive 
5-prong branching ratio (based on the CLEO branching ratios).
The systematic error due to this assumption is given in  
Table~\ref{table:syst} and was determined by assuming that the
efficiency for selecting $\fhpp$ decays is approximately 0.3
and that one third of the $\fhp$ candidates are $\fhpp$ decays.

\begin{table}
\begin{center}
\begin{tabular}{lrr} \hline
                       & $\fh$(\%) & $\fhp$(\%)  \\ \hline
Tracking               & \sxsab  & \sxsbb   \\ 
Electromagnetic energy & \sxsae  & \sxsbe  \\ 
Fit                    & \sxsad  & \sxsbd  \\ 
$\fhpp$                & \sxsag  & \sxsbg  \\ \hline
Total                  & \sxsat  & \sxsbt  \\ \hline
\end{tabular}
\end{center}
\caption{\label{table:syst} The systematic uncertainties of the $\fh$
and $\fhp$ branching ratios.}
\end{table}

The inclusive 5-prong branching, $\finc$, was determined
by applying the same selection procedure used in the exclusive measurement.
The inclusive branching ratio was determined using
\[ 
B = \frac{N-N_{\qq}-N_\tau^{bkgd}}{N_{\tau} \; (1-f)\; \epsilon \;F_{\rm B}} 
\]
where $N$ is the number of 5-track jets in the sample ($\bdat$),
$N_{\qq}$ is the number of background jets from $\qq$ events ($\bqqb$),
$N_{\tau}^{bkgd}$ is the number of background jets from tau decays ($\bttb$),
$\epsilon$ is the efficiency for selecting the signal jets ($0.556$),
and $F_{\rm B}$ is the bias factor ($\bfbb$).
The branching ratio is found to be $\binc$ where the first error 
is statistical and the second is systematic.
The systematic errors are similar to those discussed above.
The error on the efficiency includes a statistical component ($0.006$),
a component ($0.020$) associated with the uncertainty in the ratio between 
the $\fh$ and $\fhp$ branching ratios quoted here and a component ($0.030$)  
for the possible contribution from the $\fhpp$ decay (discussed below).

The inclusive branching ratio was calculated assuming that the
efficiency for selecting $\fhpp$ decays is the same as the efficiency
quoted above.
However, the efficiency for selecting $\fh$ and $\fhp$ decays was found to 
$0.58$ and $0.46$, respectively, so this assumption may not be valid.
The difference in the $\fh$ and $\fhp$ efficiencies was found to be
due to the requirement that there be five tracks in each candidate jet.
The $\fhp$ decays contain a $\pi^0$ that can produce additional tracks
by a Dalitz decay or photon conversion.
The distribution of the number of electromagnetic clusters was found
to have the same shape for the $\fh$ and $\fhp$
so the cut on this quantity is not a concern; 
this is not unexpected as tau decays are highly collimated and a 
coarse clustering algorithm is used.
The systematic error on our  efficiency ($0.030$) 
was obtained by assuming that 10\% of the inclusive branching ratio
is due to $\fhpp$ decays (based on the CLEO branching ratios)
and that the efficiency for selecting these
decays is approximately $0.30$.

\newpage
\noindent
{\large\bf Summary}

\noindent
The  branching ratios of the 
decay of the $\tau$ lepton to five charged particles 
are found to be
\[ 	B(\tauch) = \bpone   \]
\[	B(\tauneut) = \bptwo \]
where the first error is statistical and the second systematic.
These results are in good agreement with previous measurements
\cite{aleph_5prong,cleo_5prong,argus_5prong,hrs_5prong}.
These results when combined with other measurements of tau
decays to five pions are found to be consistent with the prediction of an 
isospin model \cite{sobie, rouge}.

\bigskip\bigskip\bigskip\bigskip
\appendix
\par
\noindent
{\bf Acknowledgements:}
\par
\noindent
We particularly wish to thank the SL Division for the efficient operation
of the LEP accelerator at all energies
 and for their continuing close cooperation with
our experimental group.  We thank our colleagues from CEA, DAPNIA/SPP,
CE-Saclay for their efforts over the years on the time-of-flight and trigger
systems which we continue to use.  In addition to the support staff at our own
institutions we are pleased to acknowledge the  \\
Department of Energy, USA, \\
National Science Foundation, USA, \\
Particle Physics and Astronomy Research Council, UK, \\
Natural Sciences and Engineering Research Council, Canada, \\
Israel Science Foundation, administered by the Israel
Academy of Science and Humanities, \\
Minerva Gesellschaft, \\
Benoziyo Center for High Energy Physics,\\
Japanese Ministry of Education, Science and Culture (the
Monbusho) and a grant under the Monbusho International
Science Research Program,\\
German Israeli Bi-national Science Foundation (GIF), \\
Bundesministerium f\"ur Bildung, Wissenschaft,
Forschung und Technologie, Germany, \\
National Research Council of Canada, \\
Research Corporation, USA,\\
Hungarian Foundation for Scientific Research, OTKA T-016660, 
T023793 and OTKA F-023259.\\

%

\newpage
\begin{thebibliography}{99}

\bibitem{pdg} 
        R.~M.~Barnett \etal,  Phys. Rev. {\bf D54} (1996) 1
        and 1997 off-year partial update for the 1998 edition available on 
        the PDG WWW pages (URL: http://pdg.lbl.gov/). 

\bibitem{aleph_5prong}
        ALEPH Collaboration, D.~Buskulic \etal,
        Z. Phys. {\bf C70} (1996) 579.
\bibitem{cleo_5prong}
        CLEO Collaboration, G.~Dibaut,
        Phys. Rev. Lett. {\bf 73} (1994) 934.
\bibitem{argus_5prong}
        ARGUS Collaboration, H.~Albrecht \etal, 
        Phys. Lett. {\bf B202} (1988) 149.
\bibitem{hrs_5prong}
        HRS Collaboration, B.G.~Bylsma \etal,
        Phys. Rev. {\bf D35} (1987) 2269. 

\bibitem{opaltauneutrino}
        OPAL Collaboration, K.~Ackerstaff \etal, 
        CERN-EP/98-055, 15th April 1998.

\bibitem{sobie} 
        R.J.~Sobie, Z. Phys. {\bf C69} (1995) 99.

\bibitem{rouge} 
        A.~Roug\'{e}, Z. Phys. {\bf C70} (1996) 65.

\bibitem{opaldetector} 
        OPAL Collaboration, K. Ahmet \etal,
        Nucl. Instr. and Meth. {\bf A305} (1991) 275; \\
        P.P.~Allport \etal, 
        Nucl. Instr. and Meth. {\bf A324} (1993) 34; \\
        P.P.~Allport \etal, 
        Nucl. Instr. and Meth. {\bf A346} (1994) 476.

\bibitem{opaltracker}
        M.~Hauschild \etal,  Nucl. Instr. and Meth. {\bf A314} (1992) 74; \\
        O.~Biebel \etal,  Nucl. Instr. and Meth. {\bf A323} (1992) 169; \\
        M.~Hauschild,  Nucl. Instr. and Meth. {\bf A379} (1996) 436.

\bibitem{koralz}
        S.~Jadach, B.~F.~L.~Ward and Z.~Was,
        Comp. Phys. Comm.  {\bf 79} (1994) 503.

\bibitem{tauola}
        R.~Decker, S.~Jadach, J.~H.~K\"uhn and Z.~Was,
        Comp. Phys. Comm. {\bf 76} (1993) 361.

\bibitem{gopal}
        J.~Allison \etal, Nucl. Instr. and Meth. {\bf A317} (1992) 47.

\bibitem{opal:electron}
        OPAL Collaboration, G. Alexander \etal,
        Phys. Lett. {\bf 369} (1996) 163.

\bibitem{jetset}
        T.~Sj\"{o}strand, Comp. Phys. Comm. {\bf 82} (1994) 74.

\bibitem{idncon}
        OPAL Collaboration, G. Alexander \etal,
        Z. Phys. {\bf C70} (1996) 357.

\bibitem{opal_k0_paper}
        OPAL Collaboration, R.~Akers \etal,
        Phys. Lett. {\bf 339} (1994) 278.

\end{thebibliography}
\end{document}